\documentclass[mathleft
]{an}
\usepackage{graphicx}
\usepackage{times}
\begin{document}

\Pagespan{789}{}
\Yearpublication{2006}%
\Yearsubmission{2005}%
\Month{11}%
\Volume{999}%
\Issue{88}%

\title{New Generation Stellar Physics: Asteroseismology \& Virtual Observatory}

\author{J.C.~Su\'arez\inst{1}\fnmsep\thanks{Corresponding author:
  \email{jcsuarez@iaa.es}\newline}
\and E.~Solano\inst{2,3}
\and C.~Rodrigo\inst{2,3}
\and A.~Moya\inst{3}
\and A.~Garc\'{\i}a Hern\'andez\inst{1}
}
\titlerunning{Asteroseismology \& Virtual Observatory}
\authorrunning{J.C. Su\'arez et al.}
\institute{
Instituto de Astrof\'{\i}sica de Andaluc\'{\i}a (CSIC), Rotonda de la Astronom\'{\i}a S/N,
Granada, 3004, Granada, Spain.
\and 
LAEX-CAB (INTA-CSIC), PO BOX 78, 28691 Villanueva de la Cañada, Madrid, Spain
\and 
Spanish Virtual Observatory (SVO): http://svo.cab.inta-csic.es}

\received{30 May 2005}
\accepted{11 Nov 2005}
\publonline{later}


\abstract{In the last years we have witnessed a dramatic change in the research infrastructures:
Advances in communication networks, computational resources and data storage devices are fostering
new and more efficient science. In this new scenario, the Virtual Observatory (VO) is the framework
where a new methodology for astronomical research is being built. This poster shows the natural
connection between Asteroseismology and VO. We describe the current status of a project developed
by the Spanish Virtual Observatory in which, for the first time, asteroseismic models together
with visualization tools for asteroseismology are managed within VO.}

\maketitle

\section{Introduction\label{sec:intro}}

The twentieth century has provided an important qualitative and quantitative leap in the development
of science and technology. The perfect symbiosis between the two has o\-pen\-ed new horizons in the
knowledge of nature:  from the study of the fundamental components of matter, through life sciences,
to better understand of the origin and evolution of the Universe and its components, like galaxies,
stars, planets, to name a few.

In the last decades we have witnessed a significant pro\-gress of Stellar Physics, largely due to
the
development of its great laboratory: the stellar seismology (also known as asteroseismology). Today,
several space missions have been designed to study the interior of stars and search for extra-solar
pla\-nets like MOST (Matthews et al.~1998), CoRoT (Baglin et al.~2003), or Kepler (Gilliland et
al.~2010) and some others are in preparation like PLATO (Catala 2009). 
Besides this, se\-ve\-ral precision
instruments are being used for the first time for asteroseismology (CRIRES, UVES, FOCES, FEROS,
HARPS) and others are currently in phase of development: NAHUAL (Mart\'{\i}n et al.~2005), 
GIANO (Oliva et al.~2006), CARMENES (Quirrenbach et al.~2009), etc.

Moreover, one of the great revolutions in data trans\-mi\-ssion is undoubtedly the emergence
of the
Internet and its boundless possibilities. Day after day, new Internet-based technologies such as
Web, Web 2.0, social networks, etc. come up with new useful tools. In astrophysics, this
breakthrough has its own name: the Virtual Observatory\footnote{http://ivoa.net/} (VO). Today, this
tool allows us to handle hundreds of terabytes in a single click, and access to thousands of
astronomy da\-ta\-ba\-ses interlinked through a single web portal.

We present a new generation tool for stellar physics: the fusion of Virtual Observatory
technologies with as\-te\-ro\-seis\-mo\-lo\-gy. The result is a tool conceived to easily
handle observations
and models in the framework of the large research programs like the above mentioned space missions.
We seek to offer to the community a tool from which users have access to very different models
without building interfaces, easily find models representative of the studied stars, and even
compare online global and shell physical variables  of a selection of models.

\section{The project details\label{sec:project}}

The official name of the project (the name of the tool is to be defined) is "Development of a
VO-tool for asteroseismic models". It is currently under development
within the Spa\-nish Virtual Observatory Project\footnote{http://svo.cab.inta-csic.es/} (SVO).

The staff is composed by four researchers: A. Moya, A. Garc\'{\i}a Hern\'andez, E. Solano
(project manager) and J.C. Su\'arez (principal investigator), and one technician (C. Rodrigo), from
two institutions: the Instituto de Astrof\'{\i}sica de Andaluc\'{\i}a (IAA-CSIC), and the Centro de
Astrobiolog\'{\i}a (CAB-INTA-CSIC).

Presently, around $5\,10^5$ asteroseimic models are im\-ple\-men\-ted in the tool. These have
been
computed using three different codes: 

\begin{itemize}
 \item [$\bullet$] The evolutionary code {\sc CESAM} (Morel 1997) which provides the equilibrium
                   models.
 \item [$\bullet$] The {\sc filou} code (Su\'arez 2002; Su\'arez, Goupil \& Morel 2006; Su\'arez \&
                   Goupil 2008) which provides adiabatic oscillations corrected for the effect of
                   rotation.
 \item [$\bullet$] The GraCo code (Moya et al.~2004; Moya \& Garrido 2008a).
\end{itemize}

This VO service is flexible enough to incorporate, in the future, other collections coming from
different codes and/or databases that might be of interest for other groups. 

We have implemented in the tool the following technical services

\begin{itemize}
 \item [$\bullet$] A S3 server for each
                   code\footnote{http://svo.cab.inta-csic.es/theory/sisms3}.
 \item [$\bullet$] A web application using he S3 services to find, select and analyze
                   data\footnote{http:svo.cab.inta-csic.es/theory/sisms3/}.
 \item [$\bullet$] The first \emph{data model} for asteroseismic 
                   data\footnote{http://svo.cab.inta-csic.es/theory/sisms3/concepts.php}.
\end{itemize}

\begin{figure}
   \includegraphics[width=8cm]{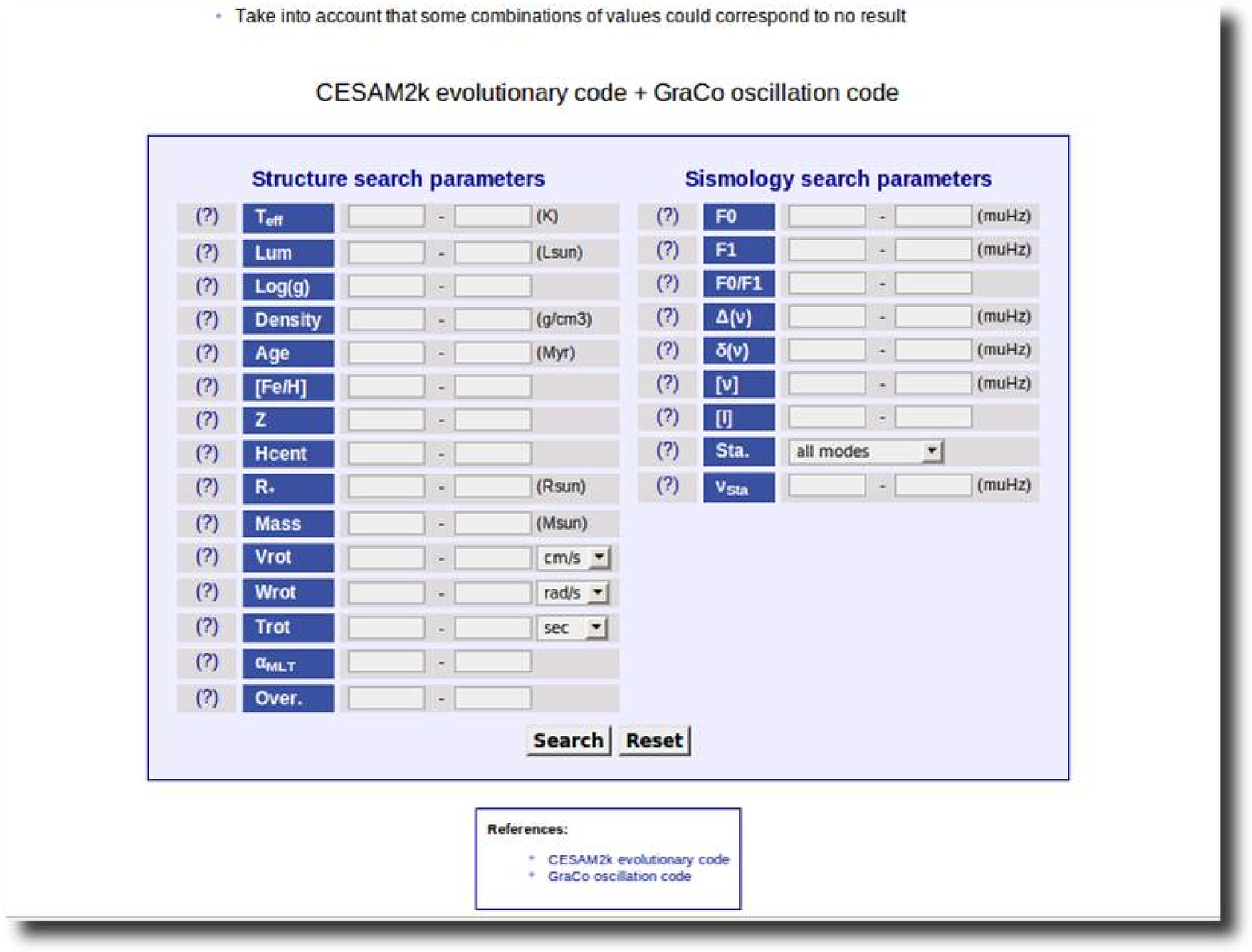}
    \caption{Snapshot of the web user interface in which the user provides input parameters
             for search queries. Global stellar parameters together with equilibrium model physical
             parameters (left column), and pulsation variables (right column).
             Prior to this interface, the user has previously selected the model databases in which
             the tool will send the user queries.}
\label{fig:input}
\end{figure}

The data model mentioned above consists in 17 stellar global properties (e.g. effective temperature,
surface gravity, luminosity, etc.), 44 star shell variables (e.g. density, pressure, temperature,
etc.), and 35 seismic properties (e.g. frequency ranges, fundamental radial mode, large
separation,
etc.). These correspond to input criteria that can be queried simultanously to the tool database
service (see Fig.~\ref{fig:input}) in order to find all the models in the databases selected by the
user. 

\begin{figure}
   \includegraphics[width=8cm]{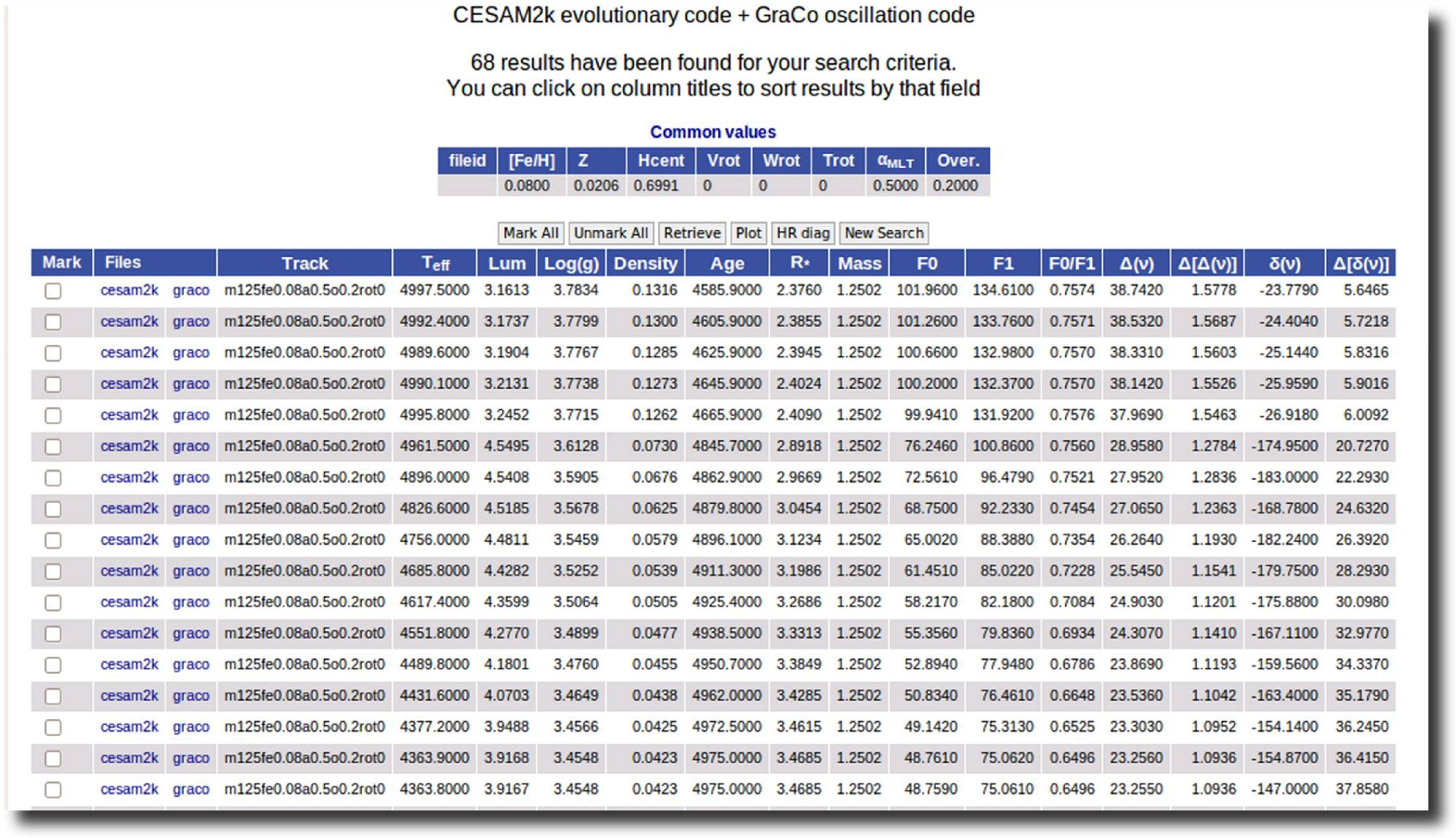}
    \caption{Snapshot with an illustration of the ouputs shown by the tool
             with the models matching all the input criteria simultanously.
             On top, a table summarizes the common parameters/variables to all
             the resulting models, and the number of \emph{valid} models, i.e. those
             matching all the user-defined input criteria simultanously.
             VOtables and complete models can be downloaded, in complete tracks
             or individually by selecting them in the results table.
             From this interface, models can be marked to be analyzed with visualization
             tools.}
\label{fig:result}
\end{figure}

Figure~\ref{fig:result} illustrates the results yield by the tool, given some inputs parameters
queried by the user. This include information on the number of \emph{valid} results, i.e. the
number of models matching the input criteria, the common values (databases might be very
heterogenous, so common physical parameters provide an idea of which variables are simultanously
shared by all the models), and the list of valid models. This list contains some useful variables,
like mass, metallicity, effective temperature, surface gravity, etc. Models can be sorted by any of
these variables. Moreover, it is possible to visualize and/or dowload each of the selected models
together with the model track they belong to\footnote{Databases are constructed by including model
evolutionary tracks. For each of the models of the tracks, the oscillation spectrum is also
provided.}.

\begin{figure}
   \includegraphics[width=8cm]{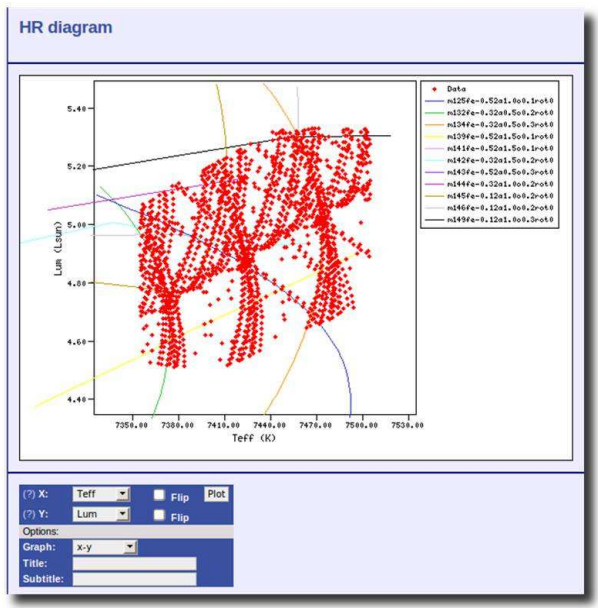}
    \caption{Snapshot with an illustration of a Herzprung-Russel diagram of the models
             matching all the input criteria simultanously. Dots correspond to the effective
             temperature and luminosity of all the valid models, and lines represent some
             of the selected evolutionary tracks.}
\label{fig:hr}
\end{figure}

\begin{figure}
   \includegraphics[width=8cm]{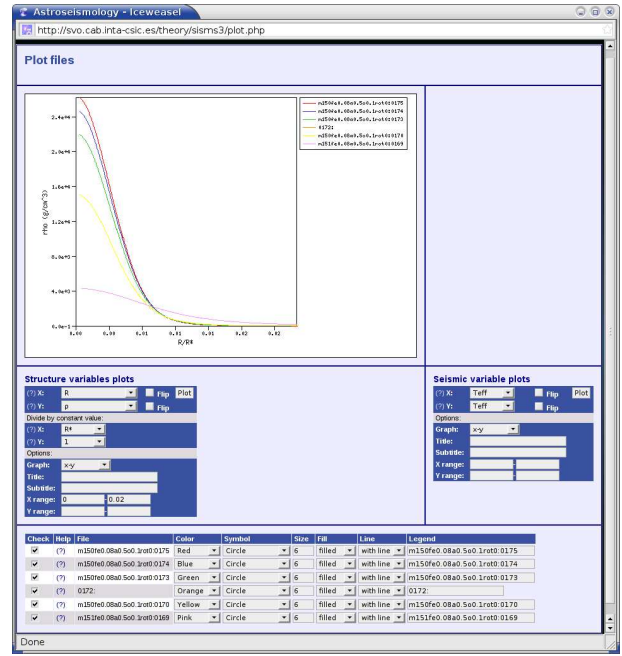}
    \caption{Snapshot with an illustration of a shell-variable type diagram of the models
             matching all the input criteria simultanously. In particular, this diagram depicts
             the variation of the density (in c.g.s.) as a function of the radial distance
             normalized to the stellar radius. 
              }
\label{fig:shell}
\end{figure}

For all the valid models yield by the tool, it is possible to make diagrams with global and shell
variables. An example of a plot with global variables is the HR diagram shown in
Fig.~\ref{fig:hr}.
Plots are flexible and allow modification of scales (e.g. for zooming and/or using a
logarithmic scale) and variables. Moreover, interactive shell variable plots are also implemented.
These consist in diagrams of in-shell variables (i.e. variables varying with the model shell, for
instance with the stellar radius), in which the user can compare the behavior of such variables for
different selected models (see e.g. Fig.~\ref{fig:shell}).

To summarize, the main characteristics of this VO tool are:
\begin{itemize}
 \item [$\bullet$] Efficiency. The VO tool queries multiple model da\-ta\-ba\-ses in seconds.
 \item [$\bullet$] Collections of models are handled easily and with user-friendly web interfaces.
 \item [$\bullet$] The only software required is a web browser with Ja\-va\-script.
 \item [$\bullet$] Tables, figures, and model collections are fully downloadable.
 \item [$\bullet$] Designed for the easy and rapid comparison of very different and heterogenous
                   models.
 \item [$\bullet$] Visualization tools are available
 \item [$\bullet$] The tool offers new scientific potential.
\end{itemize}

The posibility of managing a huge amount of models, their online comparison and visualization,
opens new possibilities in the research itself. For instance, the user of this tool will be able to
search for general properties in very different and heterogeneous databases, to make statistics, or
simply focus in the science, instead of spending a significant amount of time building interfaces,
dealing with formats, writing codes for plotting results, etc. An exemple of application is the
recently comparison (Moya et al.~2008b) of oscillation codes undertaken within the scientific
exploitation of the asteroseismic space mission \emph{CoRoT}. 

\section{Future prospects\label{sec:status}}

The present VO tool will be \emph{see the web light} during this year. This will be done in a
scientific publication with peer review. To do so, we are currently preparing a scientific case,
which is mainly based on oscillation mode statistics for intermediate-mass, main-sequence stars
(see, for instance, Garc\'{\i}a-Hern\'adez et al.~2009, Poretti et al.~2009).

Then, once the tool is published, we will initiate a contact round with research groups interested
in adapting their model databases to VO. 

From the point of view of technical implementation, it is envisaged to link directly observational
input parameters from other VO tools. This will empower the tool for the use with large scientific
programs producing lots of data, like space missions. 

In what regards the modeling, we are currently studying the possibility of linking this tool with
GRID computing-based technology, and the use of genetic algorithms.

\acknowledgements
   JCS acknowledges support from the "Instituto de Astrof\'{\i}sica de Andaluc\'{\i}a (CSIC)" by an
   "Excellence Project post-doctoral fellowship" financed by the Spanish
   "Consejerer\'{\i}a de Innovaci\'on, Ciencia y Empresa de la Junta de Andaluc\'{\i}a" under
    proyect "FQM4156-2008". JCS also acknowledges support by the Spanish "Plan Nacional del Espacio"
   under project ESP2007-65480-C02-01.


\end{document}